\documentclass[aps,prl,reprint,preprintnumbers,superscriptaddress,amsmath,amssymb,bibnotes,longbibliography]{revtex4-2}

\usepackage{graphicx}
\usepackage{dcolumn}
\usepackage{bm}
\usepackage{color}
\usepackage[colorlinks,linkcolor=blue,anchorcolor=blue,citecolor=blue,breaklinks,CJKbookmarks=True,urlcolor=blue,filecolor=blue,menucolor=blue,runcolor=blue]{hyperref}
\usepackage[utf8]{inputenc}

\begin{document}
	
	
	\title{Magnetic states of the Kondo lattice Ce$_2$PdSi$_3$ and their pressure evolution}

\author{Yanan Zhang}
\affiliation  {New Cornerstone Science Laboratory, Center for Correlated Matter and School of Physics, Zhejiang University, Hangzhou 310058, China}

\author{Zhaoyang Shan}
\affiliation  {New Cornerstone Science Laboratory, Center for Correlated Matter and School of Physics, Zhejiang University, Hangzhou 310058, China}

\author{Jiawen Zhang}
\affiliation  {New Cornerstone Science Laboratory, Center for Correlated Matter and School of Physics, Zhejiang University, Hangzhou 310058, China}

\author{Kaixin Ye}
\affiliation  {New Cornerstone Science Laboratory, Center for Correlated Matter and School of Physics, Zhejiang University, Hangzhou 310058, China}

\author{Yongjian Li}
\affiliation  {New Cornerstone Science Laboratory, Center for Correlated Matter and School of Physics, Zhejiang University, Hangzhou 310058, China}

\author{Dajun Su}
\affiliation  {New Cornerstone Science Laboratory, Center for Correlated Matter and School of Physics, Zhejiang University, Hangzhou 310058, China}

\author{Pascal Manuel}
\affiliation{ISIS Facility, STFC, Rutherford Appleton Laboratory, Chilton, Didcot, Oxfordshire OX11 0QX, United Kingdom}

\author{Dmitry Khalyavin}
\affiliation{ISIS Facility, STFC, Rutherford Appleton Laboratory, Chilton, Didcot, Oxfordshire OX11 0QX, United Kingdom}

\author{Devashibhai Adroja}
\affiliation{ISIS Facility, STFC, Rutherford Appleton Laboratory, Chilton, Didcot, Oxfordshire OX11 0QX, United Kingdom}
\affiliation{Highly Correlated Matter Research Group, Physics Department, University of Johannesburg, PO Box 524, Auckland Park 2006, South Africa}

\author{Daniel Mayoh}
\affiliation{Department of Physics, University of Warwick, Coventry CV4 7AL, United Kingdom}

\author{Geetha Balakrishnan}
\affiliation{Department of Physics, University of Warwick, Coventry CV4 7AL, United Kingdom}

\author{Yu Liu}
\affiliation  {New Cornerstone Science Laboratory, Center for Correlated Matter and School of Physics, Zhejiang University, Hangzhou 310058, China}

\author{Michael Smidman}
\email{msmidman@zju.edu.cn}
\affiliation  {New Cornerstone Science Laboratory, Center for Correlated Matter and School of Physics, Zhejiang University, Hangzhou 310058, China}

\author{Huiqiu Yuan}
\email{hqyuan@zju.edu.cn}
\affiliation  {New Cornerstone Science Laboratory, Center for Correlated Matter and School of Physics, Zhejiang University, Hangzhou 310058, China}
\affiliation  {Institute of Fundamental and Transdisciplinary Research, Zhejiang University, Hangzhou 310058, China}
\affiliation  {Institute for Advanced Study in Physics, Zhejiang University, Hangzhou 310058, China}
\affiliation  {State Key Laboratory of Silicon and Advanced Semiconductor Materials, Zhejiang University, Hangzhou 310058, China}

	\date{\today}

	\begin{abstract}
		Frustrated Kondo lattices are ideal platforms for exploring unconventional forms of quantum criticality, as well as magnetism and other emergent phases. Here we report the magnetic properties of the candidate frustrated heavy fermion compound Ce$_2$PdSi$_3$, and map their evolution upon applying magnetic fields and hydrostatic pressure. We find that at ambient pressure Ce$_2$PdSi$_3$ exhibits two distinct magnetic phase transitions, a ferromagnetic-like transition at $T_{\mathrm{M1}}=3.8$~K and an incommensurate antiferromagnetic transition at $T_{\mathrm{M2}}=2.9$~K. Upon applying pressure, $T_{\mathrm{M1}}$ is continuously suppressed and becomes undetectable above 4.2~GPa, whereas $T_{\mathrm{M2}}$ increases and remains robust up to at least 7.5~GPa. The observed pressure evolution of magnetic order in Ce$_2$PdSi$_3$ suggests the presence of competing magnetic orders, and cannot be simply encapsulated by the Doniach phase diagram, motivating further investigations for its origin, including discerning the role of geometric frustration.
	\end{abstract}
	\maketitle

	
	\section{\uppercase\expandafter{\romannumeral1}. INTRODUCTION}
	
The properties of Kondo lattice materials, constituting a periodic arrangement of $f$-electron ions that interact with the conduction electrons, can often be encapsulated by the Doniach phase diagram, which describes the evolution of the ground state upon tuning the competition between Kondo screening and the Ruderman--Kittel--Kasuya--Yosida (RKKY) interaction. Upon increasing the exchange coupling $J_{cf}$ between local spins and conduction electrons, the ground state can often be tuned from a magnetic phase to a heavy Fermi liquid upon crossing a zero-temperature quantum critical point (QCP) \cite{Doniach1977,Weng2016}.  Frustration can also act as a  tuning parameter, whereby the zero-point fluctuations can melt antiferromagnetic order even in the absence of Kondo screening \cite{Si2006,Coleman2010,balents2010N,vojta2018frustration}, and therefore acts as an additional tuning axis beyond the Doniach picture, leading to novel phenomena such as  unconventional quantum criticality \cite{Tokiwa2013,Zhao2019} and metallic spin liquid phases \cite{nakatsuji2006metallic, Tokiwa2015}.

The rare-earth-based ternary intermetallic compounds $R_2$PdSi$_3$ ($R$ = rare-earth element) serve as a platform for studying rare-earth based metallic frustrated magnetism, which crystallize in the AlB$_2$-type structure where the $R$ ions form a triangular lattice and the Pd/Si ions occupy the same crystallographic site, or ordered superstructure variants \cite{kotsanidis1990magnetic,szytula1999magnetic,Tang2011}. This series exhibits a diverse range of magnetic properties including the coexistence of spin-glass behavior and short-range order with magnetic  ordering \cite{li2003ac,Frontzek2007,paulose2003anisotropic}, complex magnetic ordering driven by long-range RKKY interactions \cite{MSmidman2019,Mukherjee2011,pecanha2026,Inosov2009,Ju2023,Paddison2022}, low-dimensional magnetism \cite{paulose2003anisotropic}, giant magnetoresistance \cite{mallik1998large,Majumda2000,Majumdar2001}, and anisotropic magnetocaloric effects \cite{sampathkumaran2000magnetocaloric,Majumda2000}. Of particular interest has been the observation of Bloch-type skyrmions in Gd$_2$PdSi$_3$  \cite{kurumaji2019}, which may be stabilized by frustrated anisotropic RKKY and dipolar interactions \cite{Okubo2012,Paddison2022,Utesov2022,dong2024fermi} or Dzyaloshinskii-Moriya interactions enabled by locally broken inversion symmetry  \cite{Moody2025}. While uniaxial compression of $\mathrm{Gd}_2\mathrm{PdSi}_3$ along the $c$ axis can expand the stability region of the skyrmion phase, suggesting a strong coupling between the lattice and topological magnetic states \cite{Spachmann2021}, hydrostatic pressure studies of $\mathrm{Gd}_2\mathrm{PdSi}_3$ remain limited, likely due to the highly localized $4f^7$ shell of $\text{Gd}^{3+}$ ions and the absence of Kondo physics \cite{Zhang2020}.
	
In contrast, the Ce-based variant Ce$_2$PdSi$_3$ has been reported to exhibit both antiferromagnetism and heavy fermion behavior, with an antiferromagnetic transition at $T_N \approx 3$~K and a Kondo temperature $T_K \approx 8$~K, suggesting  competition between the Kondo effect and magnetic interactions \cite{saha2000}.  Meanwhile powder  neutron diffraction suggested incommensurate magnetic order, with moments orientated in the basal plane \cite{szytula1999magnetic}. However, the evolution of the magnetic and electronic states with pressure has not been reported. In this work, we report detailed measurements of the magnetic properties of Ce$_2$PdSi$_3$ single crystals, including their evolution under the application of magnetic fields and hydrostatic pressure. Our results show that at ambient pressure Ce$_2$PdSi$_3$ exhibits two distinct magnetic phase transitions at $T_{\mathrm{M1}}= 3.8$ K and  $T_{\mathrm{M2}}$$ = 2.9$ K, which are confirmed by neutron diffraction measurements. Under applied pressure, $T_{\mathrm{M1}}$ is suppressed and extrapolates to zero at 4.3 GPa, while $T_{\mathrm{M2}}$ continuously increases with increasing pressure up to 7.5~GPa. These distinct pressure dependences suggest that the evolution of the magnetism cannot be straightforwardly interpreted in terms of the Doniach phase diagram.

	\section{\uppercase\expandafter{\romannumeral2}. EXPERIMENTAL METHODS}
	
	Single crystals of Ce$_2$PdSi$_3$ were grown by the optical floating zone technique \cite{Mayoh2024}. The ambient-pressure resistivity, magnetic susceptibility and  heat capacity measurements were performed using a Quantum Design Physical Property Measurement System (PPMS). Ce$_2$PdSi$_3$ crystals were polished to approximate dimensions $120 \times 80 \times 20~\mu \mathrm{m}^3$, and were then loaded into a BeCu diamond anvil cell (DAC) with an 800-$\mu$m-diameter culet. Daphne oil 7373 was used as the pressure transmitting medium. The DAC was loaded together with several small ruby balls for pressure determination at room temperature using the ruby fluorescence method \cite{Mao1986}. Electrical resistance and ac heat capacity measurements under pressure were performed in a Teslatron-PT system with an Oxford $^{3}$He refrigerator, across a temperature range of 0.3 to 300~K with a maximum applied magnetic field of 8~T. Single crystal neutron diffraction measurements were performed using the WISH diffractometer at the ISIS facility of the Rutherford Appleton Laboratory, UK \cite{wish}.

	\begin{figure}[t]
		
		\includegraphics[width=1\columnwidth]{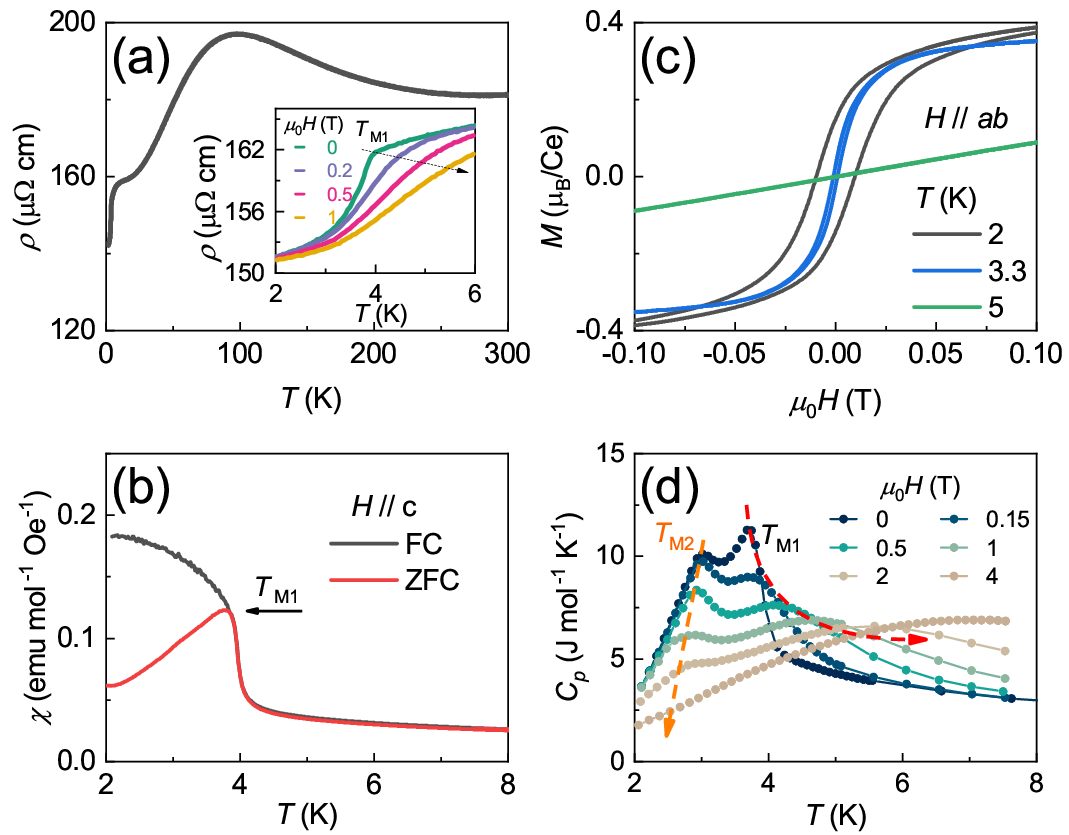}
		\caption{(Color online) (a) Temperature dependence of the electrical resistivity
			$\rho(T)$ of Ce$_2$PdSi$_3$ at ambient pressure from 300 to 2~K. The inset displays the low temperature resistivity curves measured under various magnetic fields applied along the $c$ direction, where the trend of the magnetic transition $T_{\mathrm{M1}}$ is shown. 
			(b) Low-temperature magnetic susceptibility $\chi(T)$ for $\mu_0 H = 0.1$~T with $H \parallel c$. The ZFC curve (red) was recorded on warming after cooling in zero field, whereas the FC curve (black) was recorded during field cooling from base temperature (2 K).
			(c) Isothermal magnetization $M(H)$ for $H \parallel ab$ at representative temperatures below and above $T_{\mathrm{M1}}$.
			(d) Specific heat plotted as $C_p(T)$ versus $T$ under various magnetic fields applied along the $c$ direction.}
		\label{fig1}
	\end{figure}
	
	\section{\uppercase\expandafter{\romannumeral3}. RESULTS}
	\section{{\textbf{A}}. \NoCaseChange{Magnetic ordering of Ce$_2$PdSi$_3$}}
	Figure~\ref{fig1}(a) shows the temperature dependence of the resistivity $\rho(T)$ of Ce$_2$PdSi$_3$ at ambient pressure from 300 to 2~K. Upon cooling, $\rho(T)$ first increases slowly and develops a broad maximum near $T_m \approx 100$~K. Given the relatively low Kondo temperature $T_K \approx 8$~K \cite{saha2000}, this broad feature likely originates from the Kondo scattering from the excited crystal-electric-field (CEF) levels. The inset of Fig.~\ref{fig1}(a) displays the low-temperature $\rho(T)$ under different applied magnetic fields. At zero field, an anomaly is observed around $T_{\mathrm{M1}} = 3.8$~K. $T_{\mathrm{M1}}$ increases with increasing magnetic field, which is characteristic of  ferromagnetic order. Evidence for ferromagnetism is further supported by the temperature dependence of the magnetic susceptibility $\chi(T)$ and the field dependence of the magnetization $M(H)$ shown in Fig.~\ref{fig1}(b) and (c), where a clear ZFC/FC bifurcation in $\chi(T)$ appears at $T_{M1}$ and the $M(H)$ curves display pronounced hysteresis loops between the field-increasing and field-decreasing sweeps for $T<T_{\mathrm{M1}}$. The temperature dependence of the specific heat $C_{\mathrm{p}}(T)$ under different magnetic fields is displayed in Fig.~\ref{fig1}(d). Two peaks are observed in $C_{\mathrm{p}}(T)$ with distinct field dependences: the higher-temperature peak shifts to higher temperature with increasing field, consistent with the transition at $T_{M1}$ observed in $\rho(T)$ and $\chi(T)$, whereas the lower-temperature peak moves to lower temperature with increasing field, characteristic of antiferromagnetic order. These results suggest the presence of two magnetic transitions in Ce$_2$PdSi$_3$: a possible ferromagnetic transition at $T_{\mathrm{M1}}=3.8$~K and an antiferromagnetic transition at $T_{\mathrm{M2}}=2.9$~K.
	
	\begin{figure}[t]
		\begin{center}
			\includegraphics[width=1\columnwidth]{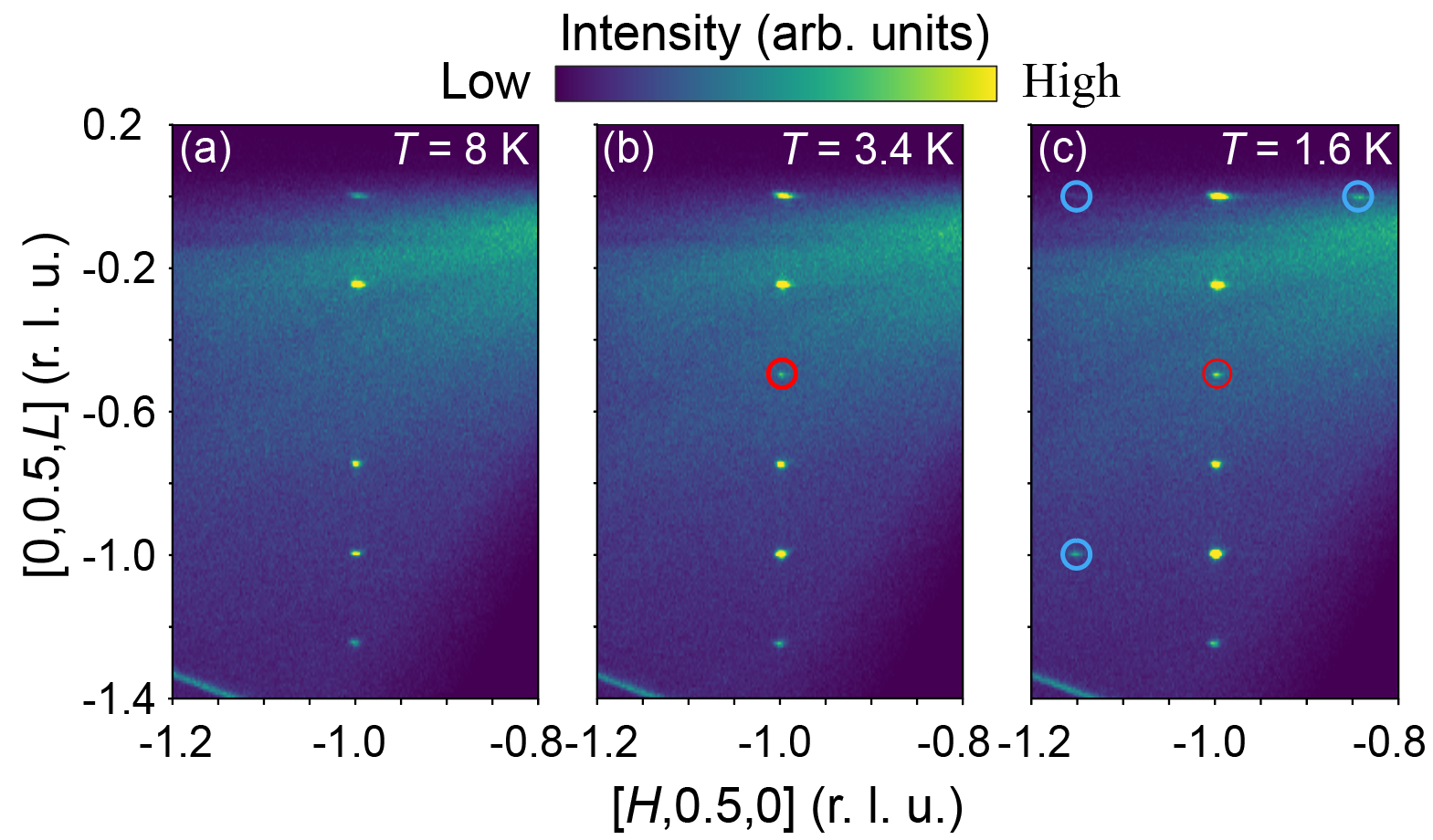}
		\end{center}
		\caption{(Color online) (a)–(c) Neutron diffraction intensity patterns of Ce$_2$PdSi$_3$ collected in the $(H,\,0.5,\,L)$ scattering plane at 8~K, 3.4~K, and 1.6~K, respectively. Magnetic Bragg peaks corresponding to the ferromagnetic and antiferromagnetic orders are highlighted by red and blue circles, respectively.}
		\label{fig2}
	\end{figure}

	To further confirm the low-temperature AFM transition, we performed single-crystal neutron diffraction at several temperatures. Figure~\ref{fig2} shows neutron diffraction intensity maps of Ce$_2$PdSi$_3$ in the $(H,\,0.5,\,L)$ scattering plane. At 8~K, several weak non-integer Bragg reflections (relative to the AlB$_2$-structure) are detected at $\mathbf{Q}=(-1,\,0.5,\,i/4)$ (integer $i\neq 2$; reciprocal-lattice units), indicating the presence of a $2\times2\times4$ structural superstructure in Ce$_2$PdSi$_3$. Such ordered superstructures are also detected in structural measurements of single crystals of other $R_2$PdSi$_3$ compounds \cite{Tang2011,pecanha2026,Paddison2022}. Upon cooling to 3.4~K below $T_{\mathrm{M1}}$, additional reflections appear at $\mathbf{Q}=(-1,\,0.5,\,-0.5)$ (red circles). The coincidence of these peaks with the structural superlattice positions indicates a propagation vector $\mathbf{k}_1=\mathbf{0}$, consistent with a ferromagnetic ordering. On further cooling to 1.6~K, additional magnetic reflections emerge (blue circles). These new peaks lie at integer-$L$ positions, implying that the associated propagation vector $\mathbf{k}_2=(0.15,\,0,\,0)$ is decoupled from the structural modulation along the $c$ axis, while it has incommensurate ordering along one direction in the $ab$ plane. The appearance of the $\mathbf{k}_2$ peaks below $T_{\mathrm{M2}}$, together with the $C_{\mathrm{p}}(T)$ data, confirms a second distinct magnetic transition corresponding to antiferromagnetic ordering.
    \begin{figure}[t]
		\begin{center}
			\includegraphics[width=0.8\columnwidth]{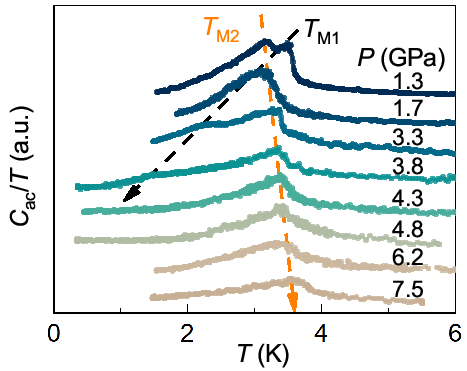}
		\end{center}
		\caption{(Color online) Temperature dependence of the ac heat-capacity coefficient $C_{\mathrm{ac}}(T)/T$ of Ce$_2$PdSi$_3$ measured at pressures between 1.3 and 7.5~GPa. Black (orange) arrows mark the trend of the FM (AFM) transitions. Note that the curves are vertically shifted for clarity.}
		\label{fig3}
	\end{figure}

	\section{{\textbf{B}}. \NoCaseChange{Magnetism of Ce$_2$PdSi$_3$ under pressure}}
	
	To determine the evolution of magnetic order under pressure, the temperature dependence of the ac specific heat $C_{\mathrm{ac}}(T)/T$ of Ce$_2$PdSi$_3$ was measured at pressures up to 7.5~GPa, as shown in Fig.~\ref{fig3}. The black and orange dashed lines indicate the opposite pressure dependences of $T_{\mathrm{M1}}$ and $T_{\mathrm{M2}}$, respectively. Specifically, $T_{\mathrm{M1}}$ is suppressed with increasing pressure, whereas $T_{\mathrm{M2}}$ is enhanced. At $P=1.7$~GPa, the two transitions nearly merge around 3~K, giving rise to a single broadened peak in $C_{\mathrm{ac}}(T)/T$. Upon further increasing pressure, $T_{\mathrm{M1}}$ decreases monotonically to 1.5 K at $P \approx 3.8$~GPa, above which it can no longer be resolved. In contrast, $T_{\mathrm{M2}}$ increases gradually with pressure, reaching 3.8~K near $P=7.5$~GPa.

\begin{figure}[t]
		\begin{center}
			\includegraphics[width=1.0\columnwidth]{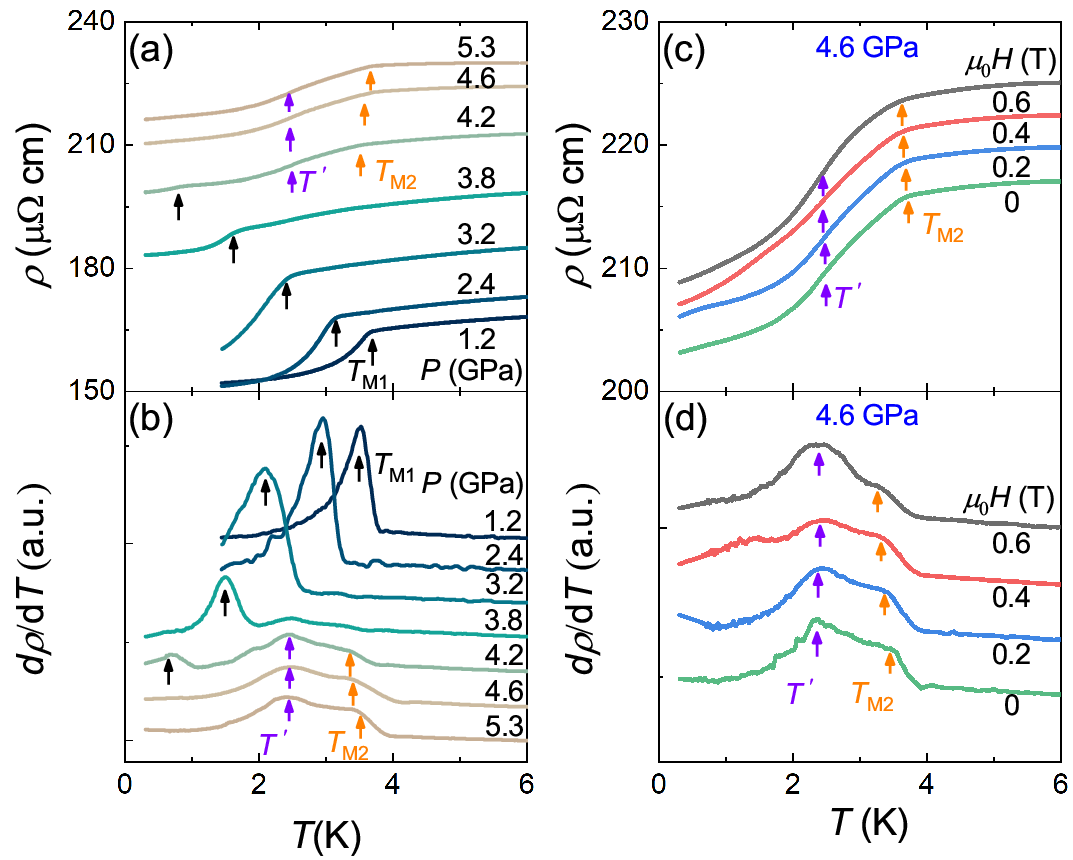}
		\end{center}
		\caption{(Color online) (a) Low-temperature resistivity $\rho(T)$ of Ce$_2$PdSi$_3$ measured at various pressures between 1.2 and 5.3~GPa. (b) Temperature dependence of the derivative of the resistivity $d\rho/dT$ under pressure. The positions of the transitions are labelled. (c) $\rho(T)$, and (d) $d\rho/dT$  at 4.6~GPa under various applied magnetic fields. Black, purple, and orange arrows mark the transition temperatures $T_{C}$, $T'$, and $T_{N}$, respectively. Note that the curves in panels (c) and (d) are vertically shifted for clarity.}
		\label{fig4}
	\end{figure}

	The resistivity $\rho(T)$ of Ce$_2$PdSi$_3$ under various pressures is shown in Fig.~\ref{fig4}. For $P<3.8$~GPa, only the ferromagnetic-like transition is observed in $\rho(T)$. The transition temperature $T_{\mathrm{M1}}$, determined from the maximum in $d\rho/dT$ (black arrows in Fig.~\ref{fig4}(b), is gradually suppressed with increasing pressure. Upon further increasing the pressure to $P=4.2$~GPa, an additional anomaly associated with the antiferromagnetic transition becomes discernible at $T_{\mathrm{M2}}=3.5$~K. The $T_{\mathrm{M2}}$ values extracted from $\rho(T)$ at higher pressure are consistent with those determined from $C_{\mathrm{ac}}(T)$. Upon cooling at 4.2~GPa, an additional anomaly is observed at $T' \approx 2.5$~K. Notably, $T'$ is nearly field independent (Fig.~\ref{fig4}(c) and (d)), and its origin remains unclear.
	
\begin{figure}[t]
		\includegraphics[angle=0,width=0.4\textwidth]{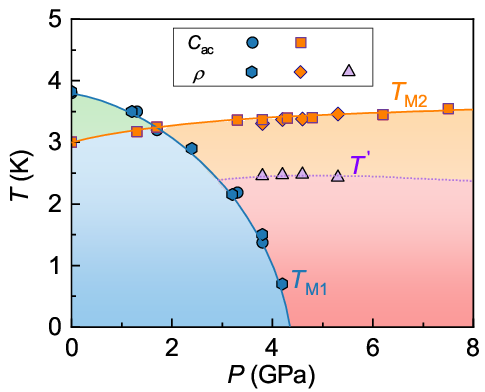}
		\vspace{-12pt} 
		\caption{(Color online)  Temperature–pressure phase diagram of Ce$_2$PdSi$_3$ in zero applied magnetic field. Blue, purple, and orange symbols denote $T_{\mathrm{M1}}$, $T'$, and $T_{\mathrm{M2}}$, respectively.}
		\vspace{-12pt}
		\label{fig5}
	\end{figure}

	The pressure--temperature phase diagram of Ce$_2$PdSi$_3$ is displayed in Fig.~\ref{fig5}, where the phase boundaries deduced from $\rho(T)$ and $C_{ac}(T)$ are highly consistent. With increasing pressure, $T_{\mathrm{M1}}$ decreases monotonically from $\sim 3.8$~K at ambient pressure to $0$~K at 4.2~GPa. In contrast, $T_{\mathrm{M2}}$ increases slowly with pressure over $0$ -- $7.5$~GPa. Moreover, a pressure-induced transition appears at $P \sim 4.2$~GPa with $T' \approx 2.4$~K, $T'$ is pressure independent and the microscopic origin remains to be clarified.

	\section{\uppercase\expandafter{\romannumeral4}. DISCUSSION AND CONCLUSIONS}
	
Our physical property characterizations and neutron diffraction measurements show the presence of two distinct magnetic transitions in Ce$_2$PdSi$_3$, namely a transition at $T_{\mathrm{M1}}$ = 3.8~K with ferromagnetic characteristics and a magnetic propagation vector $\mathbf{k}_1=\mathbf{0}$ with respect to the underlying superlattice structure, and an antiferromagnetic transition at $T_{\mathrm{M2}}$ = 2.9~K with an incommensurate in-plane propagation vector $\mathbf{k}_2=(0.15,\,0,\,0)$. Note that this propagation vector is different from that previously deduced from powder neutron diffraction measurements, where it was suggested that the incommensurate modulation was along the $c$ axis  \cite{szytula1999magnetic}. Here, the persistence of the $\mathbf{k}_1=\mathbf{0}$ magnetic peaks below $T_{\mathrm{M2}}$ shows the coexistence of the two types of magnetic order. In Ce-based intermetallics with multiple magnetic transitions, it is relatively uncommon for a ferromagnetic-like order to set in at a higher temperature than an antiferromagnetic order. Usually antiferromagnetic order develops first and a ferromagnetic component appears  at lower temperatures, as reported for CeRu$_2$Al$_2$B \cite{Bhattacharyya2016_CeRu2Al2B} and CeRu$_2$(Ge$_{1-x}$Si$_x$)$_2$ ($x=0.125$ and 0.25) \cite{Rainford2005_CeRu2GeSi2}. A similar sequence is found in Ce$_2$PdGe$_3$, which exhibits two antiferromagnetic transitions with a sizable ferromagnetic component developing in the tetragonal basal plane below the lower transition \cite{Bhattacharyya_Ce2PdGe3}. 

A microscopic coexistence of ferromagnetism and antiferromagnetism is also observed in Nd$_2$PdSi$_3$, but there both components onset at the same transition, and there is a gradual disappearance of the antiferromagnetic component with decreasing temperature \cite{MSmidman2019}. More interestingly, the incommensurate propagation vector $\mathbf{k}_2$ of Ce$_2$PdSi$_3$ is strikingly similar to the zero-field ordering wave vector of Gd$_2$PdSi$_3$ \cite{kurumaji2019}. However, as yet there is no evidence for corresponding field-induced skyrmion phases in Ce$_2$PdSi$_3$, which may be a consequence of the easy-plane single-ion anisotropy \cite{saha2000} as opposed to the isotropic Gd$^{3+}$ ions, as well as the antiferromagnetic order being buried below a possible ferromagnetic transition.

Under pressure, $T_{\mathrm{M1}}$ and $T_{\mathrm{M2}}$ exhibit different pressure dependences, with the former being suppressed by pressure and disappearing above 4.2~GPa, while $T_{\mathrm{M2}}$ gradually increases and remains robust up to at least 7.5 GPa. These different trends suggest that the evolution of the ground state cannot be simply encapsulated by competition between magnetic exchange interactions and the Kondo effect, whereby the magnetic ordering temperature first increases, and is then suppressed as  a function of a single coupling parameter $J_{cf}$. This is in contrast to Ce$_2$RhSi$_3$ where a single antiferromagnetic transition is detected that is monotonically suppressed by pressure \cite{Szlawska2009,Nakano2007}. Moreover, after the disappearance of $T_{\mathrm{M1}}$ under pressure, another anomaly $T'$ is detected below $T_{\mathrm{M2}}$, suggesting that there may be a change of the low temperature magnetic structure once the ferromagnetic transition disappears. Since above 4.2 GPa there is a transition directly from the paramagnetic to incommensurate  antiferromagnetic state, similar to Gd$_2$PdSi$_3$, it is of particular interest  to examine whether there are field-induced skyrmion or other topological spin textures in the high-pressure regime. More generally, it is important to employ additional structural probes to examine Ce$_2$PdSi$_3$  at ambient pressure to uncover detailed superlattice and magnetic structures, as well as their interplay.

	\section{ACKNOWLEDGMENTS}
	We thank Fabio Orlando for assistance with neutron diffraction experiments. Work at Zhejiang University was supported by the National Key R\&D Program of China (Grants No.~2022YFA1402200 and No.~2023YFA1406303) and the National Natural Science Foundation of China (Grants No.~W2511006, No.~12174332, No.~U23A20580, No.~12350710785, and No.~12204159). Work at the University of Warwick was funded by EPSRC, UK through Grants EP/T005963/1 and EP/N032128/1.
	
	
%

\end{document}